\documentclass[12pt,a4paper]{article}
\usepackage{pub}
\usepackage{draft}
\everymath{\displaystyle}

\usepackage[T1]{fontenc}
\usepackage{slashed}
\usepackage{caption}
\usepackage{subcaption} 
\usepackage{float}
\usepackage{empheq}
\usepackage{breqn}
\usepackage{comment}
\usepackage{xcolor}
\usepackage{color,soul}
\usepackage{bbold}
\usepackage{graphicx}
\usepackage{hyperref}
\usepackage{feynmp}
\usepackage{amsmath}
\usepackage{amsthm}
\usepackage{mathtools}

\def\[{\left[}
\def\]{\right]}
\def\({\left(}
\def\){\right)}

\preprint{}

\title{Stretched horizon, replica trick and off-shell winding condensate, and all that}

 \author[]{Indranil Halder\note{ihalder@g.harvard.edu} and}  \author[]{Daniel L. Jafferis\note{jafferis@g.harvard.edu}}

\affiliation{Jefferson Physical Laboratory, Harvard University, Cambridge, MA 02138, USA}

\abstract{$\alpha'$ corrections to near horizon dynamics of a Schwarzschild black hole in a large number of spacetime dimensions $D$ are governed by the worldsheet theory composed of the cigar CFT and the classical sigma model on the sphere at the horizon, along with a timelike-Liouville theory of central charge $26-D$. At leading order in weak string coupling, black hole thermodynamics is insensitive to the details of timelike Liouville theory. In this limit, we use the
Lewkowycz-Maldacena-trick motivated 
infinitesimally off-shell closed string worldsheet formalism in  [arxiv: 2310.02313] to calculate thermal entropy exactly in $\alpha'$. The leading term in the $\alpha'\to 0$ limit and the first stingy correction of our result are in precise agreement with the target space Callan-Myers-Perry formula.
}

\begin{document}
 
\maketitle

\section{Summary of the results}

 One of the important universal features of black hole physics is that the deep near-horizon geometry of almost all of the known supersymmetric black holes with finite-sized horizons contains a long AdS$_2$ throat \cite{Chamblin:1999tk}.  Universal features are often simpler, not pertaining to the details of a specific system. On the contrary, understanding finite energy excitations in AdS$_2$ has been a long-time challenge \cite{Maldacena:1998uz}. Only in recent years, it has been demonstrated that the key is to consider quantum gravity in near-AdS$_2$ spacetimes and discuss near extremal dynamics \cite{Almheiri:2014cka, Maldacena:2016upp, Nayak:2018qej}.
     In this setting, quantum mechanics of Sachdev-Ye-Kitaev \cite{Maldacena:2016hyu, Kourkoulou:2017zaj, DeBoer:2019yoe, Saad:2018bqo, Saad:2021rcu} and Jackiw–Teitelboim gravity \cite{Harlow:2018tqv, Saad:2019lba, Stanford:2019vob, Penington:2019kki, Heydeman:2020hhw}  are studied with tremendous success providing qualitative explanations to deep questions such as the black hole information puzzle \cite{Penington:2019npb, Almheiri:2019psf, Almheiri:2019hni, Geng:2020fxl, Balasubramanian:2022fiy, Akers:2022qdl}. Understanding stringy corrections systematically  with near AdS$_2$ boundary conditions remains a challenge.

 The content of this paper is another  universal feature of black holes: far away from extremality, for uncharged black holes in a large number of spacetime dimensions $D$, the near horizon geometry contains a long cigar throat. We pause to explain this for one of the simplest solutions to vacuum Einstein's equations - the Schwarzschild black hole. It has natural length scales set by the radius of the horizon $r_0$ and the radius of the Euclidean time circle $R=2r_0/D$. The generic quasi-normal mode of frequency  $\omega \sim 1/R$ is localized at a distance $L=(R/2) \log(D/2)$ away from the horizon \cite{Emparan:2014cia} (here we are taking large $D$ limit, keeping the angular momentum $l\sim D$ and the overtone number $n\sim 1$; in this limit the imaginary part of the frequency is sub-dominant to the real part and hence, omitted).\footnote{There are also special quasi-normal modes which are particularly light $|\omega| \sim 1/(D R ), l\sim 1$ \cite{Emparan:2014aba}, they spread over a distance $R$ away from the horizon. There has been extensive study of the effective theory of these light modes \cite{Bhattacharyya:2015dva, Dandekar:2016fvw, Bhattacharyya:2016nhn, Dandekar:2017aiv, Bhattacharyya:2017hpj, Bhattacharyya:2018iwt}. } This suggests the effect of the black hole is roughly extended in the near horizon region of length $L$, outside of which we expect empty flat space physics.  This is easy to see from the metric of the Schwarzschild black hole
 \begin{equation}
     \begin{aligned}
         & g_{\mu \nu}dx^\mu dx^\nu=\(1-\frac{r_0^{D-3}}{r^{D-3}}\) d\tau^2+ \frac{dr^2}{\(1-\frac{r_0^{D-3}}{r^{D-3}}\)}+r^2 d\Omega^2_{D-2}
     \end{aligned}
 \end{equation}
Note that for a distance of order $L$ from the horizon, i.e., $ r-r_0=L$ this reduces to flat metric in large $D$ limit. On the other hand for at distances of order $R$ from the horizon, 
the $\tau,r$ part of the metric takes the form of two-dimensional cigar geometry \cite{Emparan:2013xia} 
\begin{equation}\label{cigar}
    \begin{aligned}
        & g_{\mu \nu}dx^\mu dx^\nu\bigg|_{2d}=R^2\(d\hat{r}^2+\tanh(\hat{r})^2 d\hat{\tau}^2\), \quad \hat{\tau}=\frac{\tau}{R} , \quad  \cosh^2(\hat{r})=\(\frac{r}{r_0}\)^{D-3} 
    \end{aligned}
\end{equation}
The rescaled co-ordinates $\hat{\tau}, \hat{r}$ do not scale with $D$.\footnote{ At large D for $\hat{r}\gg1$ we can approximate the relation as
\begin{equation}
    r-r_0=\frac{2r_0}{D}\hat{r}
\end{equation}
}
Along with it we also generate a two-dimensional dilaton due to effective reduction on the sphere S$^{D-2}$. It can be easily checked that the resulting profile of the dilaton from the volume of the sphere is in the standard form of the cigar geometry as well (the precise mapping of dilaton between the flat space and the cigar geometry is a subtle question, especially the zero mode of the dilaton). 

Unlike the near extremal black hole discussed above, fortunately for the cigar geometry, we have a clear string theoretic inclusion. 
The Sigma model with cigar metric and dilaton as target space fields can be thought of as the WZW conformal field theory based on SL(2,R)$_k$/U(1) with the identification $k=R^2$ in $l_s=1$ units. The metric in (\ref{cigar}) is correct only at low temperatures that is in the $R\to \infty$ limit (in this paper, we always take $D\to \infty$ first). For finite $R$ in general the relation between $R,r_0$ will be modified from the low temperature formula $R=2r_0/D$. This low-temperature formula suggests for any large but finite value of $R$, the radius of the sphere at the horizon  $r_0 \sim D$ remains large. Moreover, as long as we are interested in $\hat{r}\ll D$ the radius of the sphere to the leading order in large $D$ is simply $r_0$.  Therefore the sigma model on S$^{D-2}$ can be treated classically at radius $r_0$ for the near horizon cigar geometry. Based on these considerations it has been proposed in \cite{Chen:2021emg} that the near horizon worldsheet theory is composed of a product of the SL(2,R)$_k$/U(1) with the classical sigma model on $S^{D-2}$ of radius $r_0$ such that the total central charge is $D$ (one can also think of it as a S-wave reduction on the sphere). The central charge constraint determines $k$ in terms of $r_0$ as follows
\begin{equation}\label{levelWZW}
    \begin{aligned}
        k=R^2=\bigg(\frac{2r_0}{D}\bigg)^2+2
    \end{aligned}
\end{equation}
It has the expected low-temperature limit mentioned above. The resulting worldsheet theory is made critical by the addition of an additional conformal field theory of central charge $26-D$ along with $bc$ ghosts. A natural choice for the additional conformal field theory might be the time-like Liouville theory. To the leading order in weak string coupling the thermodynamics of the black hole (when expressed in terms of $D$ dimensional Newton's constant $G_D$) will be unaffected by the details of the additional conformal field theory because it is independent of the temperature $r_0$ (depends only on $D$). In this paper, we will always work within this limit unless otherwise specified.\footnote{We emphasize that to the leading order in small string coupling, i.e., for spherical worldsheet theory,  the string amplitudes for external vertex operators that are independent of the the degrees of freedom of the  time-like Liouville theory get factorized between the black hole degrees of freedom and that of the time-like Liouville theory. This ensures that at the level of classical string theory, for the restricted set of observables formed entirely from the black hole degrees of freedom (and $b,c$ ghosts) the string amplitude is independent of the details of the  time-like Liouville theory except through the $D$ dimensional Newton's constant. This limit has been previously studied by the authors of \cite{Chen:2021emg}.
The argument does not apply for worldsheet of higher genus or when more general observables are considered (see section \ref{sec4}).  }  

Given that we have control over the worldsheet theory of the near horizon region one might ask to discuss the questions of black hole evaporation with stringy corrections. However, the major difficulty in discussing such questions is our lack of understanding of the origin of black hole entropy directly in the closed-string language. Using the open string description, in the light of holography \cite{Strominger:1996sh, Maldacena:1997re}, prevents us from performing a well-controlled calculation of thermodynamic qualities (and the time-dependent perturbations of that) in the strong coupling black hole region.\footnote{Other approaches based on topological strings \cite{Gopakumar:1998ii, Gopakumar:1998jq, HalderLin} are also restricted to only BPS entropy. } A step towards explaining the closed string origin of thermal entropy is taken in the recent work \cite{Halder:2023adw}. Conceptually the authors of \cite{Halder:2023adw} performed the Lewkowycz-Maldacena replica trick of the target space \cite{Lewkowycz:2013nqa}\footnote{Earlier work on the subject is due to Ryu-Takayanagi \cite{Ryu:2006bv}. For further development of the replica trick and current status see \cite{Hubeny:2007xt, Faulkner:2013ana, Engelhardt:2014gca}.} in the worldsheet language for the BTZ black hole. In the replica trick, we are required to evaluate the difference between the log of the partition function evaluated in the smooth on-shell BTZ geometry and another geometry that looks like the BTZ black hole all the way up to near the horizon, but very close to the horizon, this geometry features a conical singularity of strength $\delta$ and is therefore off-shell. The entropy is obtained by finally taking a derivative of the difference with respect to $\delta$. On the worldsheet, the on-shell BTZ geometry is presented by a linear dilaton coupled with $\beta\gamma$ system deformed by marginal $\pm1$ winding operators that preserve the U(1)$\times$U(1) isometry of the thermal background \cite{Berkooz:2007fe, Jafferis:2021ywg, Halder:2022ykw, Halder:2023nlp}. The conically singular off-shell geometry is obtained by replacing the winding operators mentioned previously with the one obtained from them by formally changing the temperature a little, essentially they carry fractional winding $\pm(1-\delta)$, preserve U(1)$\times$U(1) symmetry and hvae weight (1,1) in the unmodified free linear dilaton and $\beta \gamma$ system. In this paper, we generalize this method to the cigar geometry when we represent it as the Sine-Liouville theory using the FZZ duality. Sine-Liouville theory contains a linear dilaton for the radial direction of the cigar and a compact boson for the compact Euclidean time circle deformed by marginal winding operators on the Euclidean time circle of winding $\pm1$.
Only subtlety in following the procedure of \cite{Halder:2023adw} here is related to the fact that in cigar geometry changing the asymptotic temperature, i.e., changing $R$ also changes the slope of the linear dilaton in the Sine-Liouville language. We argue that the correct procedure is to allow for the temperature on a cut-off surface $\hat{r}_c$, the slope of the linear dilaton, and the value of the dilaton at the tip of the cigar $\hat{r}=0$ to vary while keeping the asymptotic value of the dilaton on a cut-off surface $\hat{r}_c$  fixed. Again this leads to the winding condensate of fractional winding when looked at from the un-modified theory of linear dilaton and the compact boson. From the details of the calculation, it turns out that the variation required for the entropy comes only from the change in the winding operators, the variation of the dilaton on the tip of the cigar does not contribute. The final formula for the contribution of the near horizon geometry to the entropy of the large $D$ black hole takes the following form
\begin{equation}\label{NHEntropy}
\begin{aligned}
    & S_{NH}=S_{NH,0}  \frac{(1-3 (2\pi T)^2)}{\left(1-2 (2\pi T)^2\right)^{\frac{5}{2}}}, \quad T=\frac{1}{2\pi R} \\
    & S_{NH,0}=a_{D-2}\(\frac{1}{(2\pi T)^2}-2\)^{\frac{D-2}{2}}, \quad  a_{D-2}=\frac{A_{D-2} }{4G_D}\(\frac{D}{2}\)^{D-2}
\end{aligned}
\end{equation}
Here $A_{D-2}$ is the area of unit radius $S^{D-2}$. At low temperatures, i.e., $T\to 0$ limit, the leading and sub-leading corrections obtained from this formula are in agreement with the calculation done by the target space effective action (up to four derivative order) of  Callan-Myers-Perry \cite{Callan:1988hs}. This effective action is the same as that of Bosonic string theory in $D=26$ and then just replacing Newton's constant $G_{26}\to G_D$. On the other hand, this formula shows the stringy correction to the cigar entropy $S_{NH}/S_{NH,0}$ keeps growing with increasing temperature until very close to the Hagedorn temperature of flat space string theory at $R=2$ \cite{Gross:2021gsj, Chen:2021dsw}. At this point, a winding tachyon develops in the asymptotically flat space and the contribution of the near-horizon cigar geometry is no longer the major contribution to the entropy of the large $D$  Schwarzschild black hole (in this paper, we do not evaluate the contribution of the winding tachyon in the asymptotic flat space). The entropy of the near horizon region decreases as we keep increasing the temperature and vanishes\footnote{This just means that the entropy stops being order $g_s^{-2}$.  } at the Hagedorn temperature $R=\sqrt{3}$ of empty linear dilaton spacetime. Near this temperature, our formula for the entropy reduces to the Hagedorn form. 
 
The paper is organized as follows. In section \ref{sec2}, we review the target space effective action and the solutions of Callan-Myers-Perry \cite{Callan:1988hs} keeping in mind the large $D$ limit, also we review the map between the large $D$ and cigar fields following Chen-Maldacena \cite{Chen:2021emg}. Along the way, we establish a precise map of string coupling (zero mode of the dilaton) and explain its physical meaning in terms of `stretched horizon'. In section \ref{sec3} we discuss the worldsheet theory of the near horizon geometry and perform the replica trick calculation using the winding condensate description. In section \ref{sec4} we discuss the strings in leading order in large $D$ at ultra-low temperatures $R \sim \sqrt{D}$ and show that for a certain special set of vertex operators involving modes of the $D$ dimensional spacetime and  the time-like Liouville theory, the resulting worldsheet theory enjoys remarkable simplification relating to Virasoro minimal string \cite{Collier:2023cyw}.

\section{Stringy Schwarzschild black hole in large D}\label{sec2}

\subsection{Corrections to the metric and the dilaton}

In this section, we discuss $\alpha'$ corrections to the Schwarzschild black hole in large spacetime dimension $D$. An attempt to define bosonic string theory in large dimensions necessarily runs into trouble due to the central charge constraint on the matter part of the worldsheet conformal field theory. As a result, one might think that any putative definition would be sensitive to the additional non-unitary theory that we need to add along with the sigma model on $D$ spacetime dimensions. However, if we are looking at the gravitational amplitudes only on the $D$ dimensional spacetime at leading order in weak string coupling the only dependence on the additional non-unitary theory would enter via the $D$ dimensional Newton's constant $G_D$, and therefore in this limit of weak gravitational coupling, we expect the existence of a universal set of stringy corrections to gravitational dynamics in large $D$. One way to define such corrections would be to take the target space effective action of bosonic strings in $D=26$ and just replace $G_{26}\to G_D$. In fact, this is exactly what is done by  Callan, Myers, and Perry \cite{Callan:1988hs}: stringy corrections to gravitational dynamics in large $D$ are defined by the following effective action in string frame\footnote{We are working in the same convention as that of Polchinski for the metric and dilation. }
\begin{equation}\label{SFeff}
    \begin{aligned}
        \frac{1}{2 (\alpha')^{\frac{D-2}{2}}}\int d^Dx \sqrt{-g} e^{-2\phi}\bigg(R+4 (\nabla \phi)^2+\frac{1}{4}\alpha' R_{\mu \nu \delta \sigma}R^{\mu \nu \delta \sigma}+O((\alpha')^{2}) \bigg)
    \end{aligned}
\end{equation}
Now we turn to focus on the Schwarzschild black hole and review the first $\alpha'$ correction as obtained by the effective action above keeping in mind the large $D$ limit. For the entire paper, we will restrict ourselves to the leading order in this limit. 
The Schwarzschild black hole solution in string frame is given by the following anstz \footnote{The mapping to \cite{Callan:1988hs} is given by
\begin{equation}
\lambda=\frac{1}{2}\alpha',    f_1=\mu, g_1=\epsilon, \phi_0=\varphi, r_0=\omega= \kappa^{\frac{1}{D-3}}
\end{equation}}
\begin{equation}\label{CMP}
    \begin{aligned}
       & g_{\mu \nu}dx^\mu dx^\nu=f(r)^2 d\tau^2+g(r)^2 dr^2+r^2 d\Omega^2_{D-2}\\
       & f(r)=f_0(r)(1+\frac{\alpha'}{2} f_1(r)),\quad g(r)=g_0(r)(1+\frac{\alpha'}{2} g_1(r)), \quad \phi(r)=\phi_0(r)+\frac{\alpha'}{2} \phi_1(r)\\
       & f_0(r)^2= g_0(r)^{-2}=1-\frac{r_0^{D-3}}{r^{D-3}}, \quad 2\pi  e^{-2\phi_0(r)}=\frac{(\alpha')^{\frac{D-2}{2}}}{4G_D}\\
       &  \phi'_1(r)=\frac{(D-2)^2(D-3)}{4}\frac{r_0^{D-5}(r^{D-1}-r_0^{D-1})}{r^D(r^{D-3}-r_0^{D-3})}\\
       & g_1'(r)+\frac{(D-3)r^{D-4}}{r^{D-3}-r_0^{D-3}}g_1(r)=-\phi_1'(r)-\frac{r}{D-2}\phi_1''(r)-\frac{(D-1)(D-3)r_0^{2(D-3)}}{2(r^{D-3}-r_0^{D-3})r^D}\\
       & f_1(r)=-g_1(r)+\frac{2}{D-2}(\phi_1(r)-r\phi_1'(r))
    \end{aligned}
\end{equation}
We have chosen conventions such that the horizon is at $r=r_0$ even after the stringy correction is taken into account. 
From now on we set $\alpha'\equiv l_s^2=1$.
In the paper, we are interested only in leading orders in large $D$ limit keeping the asymptotic string coupling $g_s=e^{\phi_0}$ and `level'
\begin{equation}\label{level}
    \begin{aligned}
        k=\bigg(\frac{2r_0}{D}\bigg)^2+2
    \end{aligned}
\end{equation}
fixed. This corresponds to a black hole with a very large horizon $r_0\sim D$ (we will see as we increase temperature we will need more and more stringy corrections governed by the $1/k$ corrections).  As explained in the introduction we expect universal behavior in the near horizon region i.e, on distance scales $\hat{r}$ of order one in string units \cite{Emparan:2013moa, Emparan:2013xia}
\begin{equation}\label{cigarR}
    \cosh^2(\hat{r})=\(\frac{r}{r_0}\)^{D-3}
\end{equation}
The differential equations in (\ref{CMP}) are to be supplemented with suitable asymptotically flat boundary conditions.
For us, it will be enough to use the following boundary conditions for the stringy corrections to the fields
\begin{equation}\label{CMPbc}
   \phi_1(\infty)= f_1(\infty)=g_1(\infty)=0
\end{equation}
For $r>r_0$, we simplify the differential equations in (\ref{CMP}) as follows
\begin{equation}\label{phi1}
    \begin{aligned}
         \phi'_1(r)=\frac{D^3}{4}\frac{r_0^{D-5}(r^{D-1})}{r^D(r^{D-3})} \implies \phi_1(r)=-\frac{D^2}{4r_0^2}\frac{r_0^{D-3}}{r^{D-3}} =-\frac{1}{(k-2)\cosh^2(\hat{r})}
    \end{aligned}
\end{equation}
\begin{equation}\label{g1}
\begin{aligned}
    & g_1'(r)+\frac{Dr^{D-4}}{r^{D-3}}g_1(r)=-\phi_1'(r)-\frac{r}{D}\phi_1''(r)-\frac{D^2 r_0^{2(D-3)}}{2(r^{(D-3)+D})}\\
    & g_1'(r)+\frac{D}{r}g_1(r)=-\frac{1}{D}\frac{D^3}{2r_0^3}\frac{ r_0^{2D-3}}{r^{2D-3}}\implies g_1(r)\sim \frac{1}{D}\frac{D^2}{r_0^2}\frac{ r_0^{2D-4}}{r^{2D-4}}
\end{aligned}
\end{equation}
\begin{equation}
\begin{aligned}\label{f1}
    & f_1(r)=\frac{2}{D}(\phi_1(r)-r\phi_1'(r))=-\frac{D^2}{2r_0^2}\frac{r_0^{D-3}}{r^{D-3}}=-\frac{2}{(k-2)\cosh^2(\hat{r})}
\end{aligned}
\end{equation}
This shows to the leading order in large $D$ we are interested in, we can set $g_1(r)=0$. This difference between `gravitational mass' correction $f_1$ and `inertial mass' correction $g_1$ exists because we are working in string frame, in Einstein frame they are the same. 
Combining these results we get the following expression for the target space fields to the leading order in large $D$ with the first stringy correction\footnote{For II strings similar solutions are available in \cite{Chen:2021qrz}.}
\begin{equation}\label{largeDfields}
    \begin{aligned}
        & \phi=\phi_0-\frac{1}{2(k-2)\cosh^2(\hat{r})}+O\(\frac{1}{k^2},\frac{1}{D}\)\\
	& g_{\mu \nu}dx^\mu dx^\nu\bigg|_{2d}= k\(\(1-\frac{2}{k}\) d\hat{r}^2+\tanh^2(\hat{r}) \(1-\frac{2}{(k-2)\cosh^2(\hat{r})}\) d\hat{\tau}^{2}\)+O\(\frac{1}{k},\frac{1}{D}\)\\
    \end{aligned}
\end{equation}
Here we have conveniently defined 
\begin{equation}\label{cigarT}
    \hat{\tau}=\frac{\tau}{\sqrt{k}}
\end{equation}

\subsection{Corrections to thermodynamics}

To discuss the thermodynamics of the black hole, one is required to go to Einstein frame. This involves an $\alpha'$ dependent field re-definition \cite{Callan:1986jb}. We won't review the details of it here, but simply quote the result from  Callan-Myers-Perry \cite{Callan:1988hs}\footnote{Note that the formula in Callan-Myers-Perry (see for instance equation (2.26), (3.1) of their paper) is written in terms  of Einstein frame quantity $$\omega_I=\omega\(1+\frac{(D-3)(D-4)}{4\omega^2}\)^{\frac{1}{D-3}}, \quad \omega =r_0$$}
\begin{equation}\label{largeDth}
    \begin{aligned}
        & S=\frac{A_{D-2}r_0^{D-2}}{4G_D}\bigg(1+ \frac{D^2}{2 r_0^2} \bigg)+O\((\alpha')^{2}, \frac{1}{D}\)\\
        & M = D\frac{A_{D-2}r_0^{D-3}}{16 \pi G_D}\bigg(1+ \frac{D^2}{4 r_0^2} \bigg)+O\((\alpha')^{2}, \frac{1}{D}\) 
    \end{aligned}
\end{equation}
Here $S, M$ are respectively the entropy and mass of the black hole. The temperature is calculated from the thermodynamic relation
\begin{equation}
    \begin{aligned}
        \frac{1}{T}=\frac{\partial S}{\partial M}=2\pi \sqrt{k}\(1+O\(\frac{1}{k^2},\frac{1}{D}\)\)
    \end{aligned}
\end{equation}
This suggests after we take a large $D$ limit,  stringy corrections would be important at high temperatures, i.e., at small values of $k$ limit. In other words, we can trust expressions in (\ref{largeDfields}) only up to the first sub-leading order in large $k$ limit. It was observed in \cite{Chen:2021emg} that the leading $1/k$ correction (first we take $D$ large and then take $k$ large) in (\ref{largeDfields}) matches the leading $1/k$ correction of the two-dimensional cigar geometry in `standard scheme' (written in string frame) \cite{Kazakov:2001pj} (the leading order solution was analyzed in \cite{Witten:1991yr, Elitzur:1990ubs, Mandal:1991tz}) \footnote{We have following mapping of variables with \cite{Tseytlin:1993df}
\begin{equation}
    n=\frac{1}{\sqrt{k-2}},\quad x=\sqrt{k-2} \hat{r}, \quad \theta=\sqrt{k-2} \hat{\tau}
\end{equation}
}
\begin{equation}\label{cigarS}
    \begin{aligned}
        & \Phi=\Phi_0-\frac{1}{4}\log\(1-\frac{2}{k}\)-\log(\cosh(\hat{r}))-\frac{1}{4}\log \(1+\frac{2}{k-2}\frac{1}{\cosh^2(\hat{r})}\)\\
        & G_{\mu\nu}dx^\mu dx^\nu=(k-2)\(d\hat{r}^2+a(\hat{r})^2d\hat{\tau}^2\), \quad a(\hat{r})=\frac{1}{\(\coth^2(\hat{r})-\frac{2}{k}\)^{\frac{1}{2}}}
    \end{aligned}
\end{equation}
provided we identify the `two derivative'\footnote{This entropy formula is not suitable for comparison to the large D black hole because of the lack of knowledge of how exactly the cigar is embedded in the large spacetime. 
To reproduce the thermodynamics of the large D black hole from the cigar geometry  we need to understand the $\alpha'$ dependence of the cut-off surface at $r_c-r_0 \sim L$ where the cigar geometry joins the large D black hole. It's not well-understood which higher derivative boundary terms to add to the cigar action to make the on-shell action well-defined (also the proper background dependence of the boundary terms for recovering thermodynamics of the large D black hole at a finite radial cut-off is unknown). We leave such questions for the future work.} entropy of the stringy cigar (see equation (3.26) in \cite{Kazakov:2001pj} for its expression) with the horizon area of the string-corrected Schwarzschild black hole in $D$ dimensional Planck units and then extrapolate it to generic radial distance\footnote{A alternative perspective towards this equation is that we can take this as the defining relation between $r,\hat{r}$ co-ordinates once we fix a convention for identifying the string coupling between the flat space and the cigar geometry. Instead, here we identified the radial coordinates by (\ref{cigarR}), and that leads to this natural identification of the string coupling. }
\begin{equation}\label{Dmap}
    \begin{aligned}
     & 2\pi  e^{-2\Phi}\(1-\frac{2}{k}\)=2\pi  e^{-2\phi} A_{D-2}r^{D-2}, \quad G_{\mu \nu}=g_{\mu \nu}\\
    \end{aligned}
\end{equation}
An earlier discussion of the formula in (\ref{cigarS}) is available in \cite{Dijkgraaf:1991ba, Tseytlin:1993df} where it is advocated that this formula is exact to all orders in $1/k$ under the assumption that the $\beta$ function equation of the bosonic string tachyon is one-loop exact and the winding tachyon does not modify the geometry. In this paper, we will work under this assumption.
The factor $(1-2/k)$ in (\ref{Dmap}) does not enter in the comparison of radial profile between $\Phi$ and $\phi$, it only modifies the mapping of $\phi_0$ to $\Phi_0$.
 The way we defined the dilaton in cigar geometry is standard - $\Phi_0$ has the interpretation of the dilaton at the tip of the cigar.
\begin{equation}\label{cDilaton}
    \begin{aligned}
        \Phi(\hat{r}=0)=\Phi_0, \quad \Phi(\hat{r}=\infty)=\Phi_0-\frac{1}{4}\log\(1-\frac{2}{k}\)-\hat{r}\equiv \Phi_{\infty}-\hat{r}
    \end{aligned}
\end{equation}
As we approach asymptotic infinity the string coupling becomes weak. Now we turn to explain the formula in (\ref{Dmap}) in more detail: far away from the horizon, string frame metric takes the following form
\begin{equation}\label{CMP}
    \begin{aligned}
       & g_{\mu \nu}dx^\mu dx^\nu=\(1-\frac{r_0^{D-3}}{r^{D-3}}-\frac{D^2}{2r_0^2}\frac{r_0^{D-3}}{r^{D-3}}\) d\tau^2+\dots\\
    \end{aligned}
\end{equation}
Apparently from this metric it looks as if the horizon is `stretched' up to $r_{SH}$
\begin{equation}\label{apprentH}
    r_{SH}^{D-3}=r_0^{D-3}\(1+2\frac{D^2}{4r_0^2}\)=r_0^{D-3}\(1+\frac{2}{k-2}\) \implies r_{SH}=r_0+\frac{2r_0}{D} \(\frac{1}{k-2} \)
\end{equation}

\begin{figure}[ht]
	   \centering
	   \includegraphics[width=0.8\textwidth]{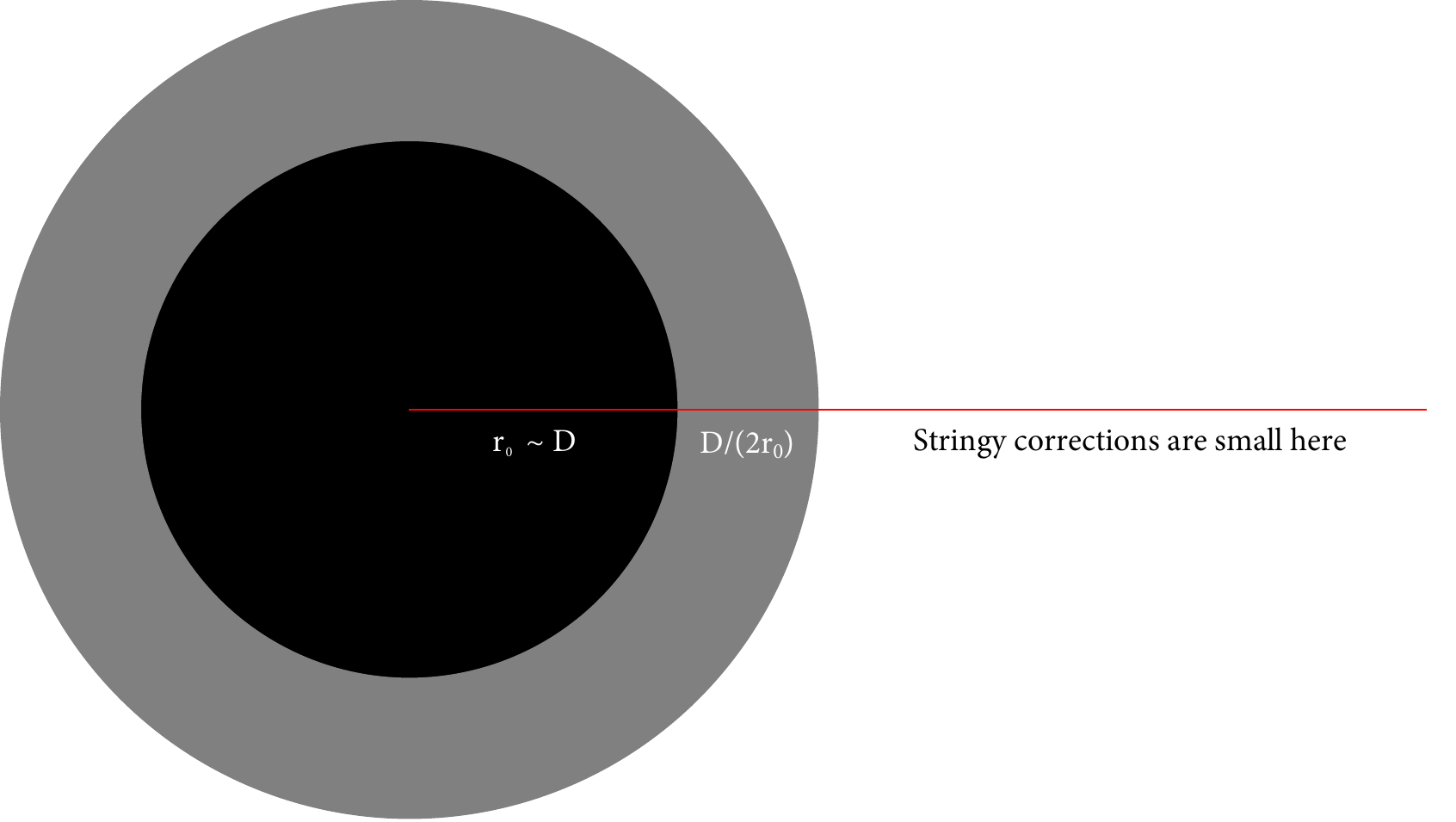}
    \vspace{0.5cm}
\caption{As we approach the horizon at $r=r_0$ of the black hole, the local temperature increases and at a location $D/(2r_0)$ away from the horizon  it reaches the Hagedorn temperature of the empty space (we are working in string frame). This leads to a tachyonic mode which winds around the Euclidean time circle. When we treat the winding tachyon in a probe approxiamtion around the black hole geometry it extends up to an additional distance of $D/(2r_0)$ (in grey) outside the horizon at $r=r_0$ (in black). Roughly speaking, stringy effects ends at the stretched horizon at $r=r_0+D/(2r_0)$. }
 \label{fig1}
\end{figure}

 As we go closer and closer to the Hagedorn transition at $k=4$ the correction term $2/(k-2)\to 1$ becomes the same in magnitude as the unmodified value. In this limit, the back reaction of the winding tachyon on the metric and the dilaton is large. Even at a generic value of $k$, we need to consider the effect of the winding tachyon that we ignored so far. The `stretched' horizon\footnote{We borrowed the terminology of `stretched' horizon from \cite{Sen:1995in, Sen:2004dp}.} in string frame extends up to $\hat{r}=1/(k-2)$ also follows from the fact that in string frame the vertex operators of the winding tachyon near asymptotic infinity decay as (to be discussed in detail later)
\begin{equation}
    W^\pm \sim e^{-(k-2)\hat{r}}
\end{equation}
As seen from asymptotic infinity, distance up to $\hat{r}=1/(k-2)$ is filled with the winding tachyon and there the back-reaction is not negligible. Outside the `stretched' horizon we can ignore the effect of the winding tachyon and previous analysis is applicable.  Note that, we can rewrite the dilation equation in (\ref{Dmap}) to leading order in large $D$ in terms of the area  of the `stretched' horizon simply as
\begin{equation}\label{Dmap0}
    \begin{aligned}
     & 2\pi  e^{-2\Phi}=2\pi  e^{-2\phi} A_{D-2}r_{SH}^{D-2} \cosh^2(\hat{r})
    \end{aligned}
\end{equation}
\textit{The dilaton at the tip of the cigar is obtained by the dimensional reduction of the $D$ dimensional dilaton at the `stretched' horizon when the back-reaction of the winding condensate is taken into account}. For related discussion of tachyon condensation and the role of `stretched' horizon see \cite{Adams:2005rb, McGreevy:2005ci, Horowitz:2006mr, Silverstein:2006tm}.\footnote{We thank Zohar Komargodski for mentioning this to us. }

\section{Classical worldsheet in large D}\label{sec3}

In this section, we define the worldsheet theory governing the near horizon dynamics following \cite{ Chen:2021emg}. And then proceed to repeat the replica trick calculation of \cite{Halder:2023adw} in this near horizon geometry. To do so we find it helpful to first point out that the BTZ calculation and the cigar calculation are intimately related through a sequence of two T dualities \cite{Berkooz:2007fe}.   Along the way we generalize and make the discussion of zero modes in \cite{Halder:2023adw} more precise. Finally, we discuss the off-shell background need for the cigar geometry carefully and compute its entropy. 

\subsection{Exact worldsheet description at leading order in large D}

Form the discussion of the previous section, it sounds very reasonable that the worldsheet theory capturing the near horizon dynamics $\hat{r}\sim 1$ is composed of SL(2,R)$_k$/U(1) $\times$ S$^{D-2}_{r_0} \times$time-like Liouville theory of central charge $26-D$  \cite{ Chen:2021emg}. The theory on the sphere of radius $r_0$ is the sigma model at leading order in the $\alpha'$ since $r_0\sim D$. $r_0, k$ are related such that the central charge of the first two factors is exactly $D$ (here we work in units $\alpha'=1$)\cite{Fradkin:1984pq}
\begin{equation}
    \begin{aligned}
     c_{\text{SL(2,R)$_k$/U(1)}}+ c_{\text{S$^{D-2}_{r_0}$}}=D\implies   k-2=\bigg(\frac{2r_0}{D}\bigg)^2
    \end{aligned}
\end{equation}
Timelike Liouville theory makes the theory critical. Since the time-like Liouville theory is independent of $k$, for the purpose of $k$ dependence of thermal entropy we can absorb this factor in the effective Newton's constant in D dimensions. The linear dilaton $\Phi$ in the cigar arises from the dimensional reduction on the sphere and therefore it is natural to define the dilaton to be given by 
\begin{equation}
    \begin{aligned}
      2\pi  e^{-2\Phi}\(1-\frac{2}{k}\)=2\pi  e^{-2\phi} A_{D-2}r_0^{D-2}\cosh^2(\hat{r})
    \end{aligned}
\end{equation}
$A_{D-2}$ is the area of unit $D-2$ sphere. This theory is to be considered only at leading order in large $D$ and for dynamics near the tip of the cigar.

\subsection{Thermal entropy form the expectation value of the area operator}

\subsubsection{Cigar geometry from the BTZ  black hole}

T duality allows us the write down locally the BTZ worldsheet theory in terms of cigar theory as we explain below. For a conceptually similar discussion in the case of thermal AdS see \cite{Berkooz:2007fe}.

Consider the worldsheet given by\footnote{Our convention is that of \cite{Halder:2023nlp}. The convention of $B$ is different from that of Polchinski's by an additional minus sign.}
\begin{equation}\label{sigma_model}
	\begin{aligned}
		& S_{cl}=\frac{1}{2 \pi}\int 2 d^2 \sigma \( \(G_{\mu \nu}+B_{\mu \nu}\) \(\partial \hat{x}^\mu \bar{\partial} \hat{x}^\nu \)+\frac{1}{4}\mathcal{R} \Phi\)\\
	\end{aligned}
\end{equation}
We break the spacetime coordinates into $\mu=\(0,a\)$, and assume that the fields are independent of the compact non-contractible coordinate $\hat{x}^0\sim \hat{x}^0+2\pi$. We can T-dualize the translation in $\hat{x}^0$ to obtain (up to one loop in $\alpha'$)\cite{Alvarez:1994dn, Giveon:1994fu}
\be\label{Tduality}
\begin{aligned}
    G'_{00}=&{1\over G_{00}}, \qquad G'_{0a}={B_{0a}\over G_{00}}, \qquad
G'_{ab}=G_{ab}-{G_{a0}G_{0b}+B_{a0}B_{0b}\over G_{00}} \nonumber\\
B'_{0a}=&{G_{0a}\over G_{00}},\qquad
B'_{ab}=B_{ab}-{G_{a0}B_{0b}+B_{a0}G_{0b}\over G_{00}}, \qquad \Phi'=\Phi-\frac{1}{2}\log G_{00}
\end{aligned}
\ee
The new fields refer to the T dual co-ordinate $(\hat{x}^{0})'\sim (\hat{x}^{0})'+2\pi$ and $\hat{x}^{a}$.

Now we will apply these general results to the special case of BTZ black hole.
At leading order in large $\sqrt{k}=l_{AdS}/l_s, l_s=1$, the sigma model fields for the Lorentzian BTZ black hole are given by
\begin{equation}\label{EBTZfields}
	\begin{aligned}
 & \Phi=\log(g_{s,3})\\
		& ds^2=k(d\hat{r}^2+ (2\pi/ \beta)^2 \cosh^2(\hat{r})d\hat{\theta}^2-\sinh^2(\hat{r})d\hat{t}^2)\\
		& B=  k (2\pi /\beta)\sinh^2 (\hat{r})  \ (d\hat{\theta} \otimes d \hat{t}-d\hat{t} \otimes d\hat{\theta})\\
	\end{aligned}
\end{equation}
Here $\beta$ is the inverse temperature of the BTZ black hole measured in AdS units.
Angular co-ordinate is periodic $\hat{\theta} \sim \hat{\theta}  + 2\pi$ (it is clear that the analytic continuation to the Euclidean signature makes the $B$ field imaginary, with this imaginary $B$ field Euclidean worldsheet CFT is well-defined). Since the $\hat{\theta}$ direction is non-contractible we can T-dualize it using the standard rules given above. 
\begin{equation}\label{T1}
	\begin{aligned}
 & \Phi'=\log(g_{s,3})-\frac{1}{2}\log\(k(2\pi \beta)^2 \cosh^2(\hat{r}) \)\\
		& ds^{'2}=k( d\hat{r}^2-\tanh^2(\hat{r})d\hat{t}^{'2}+d\hat{\chi}^{'2})\\
		& B'= 0\\
	\end{aligned}
\end{equation}
For notational convenience, we defined new co-ordinates according to
\begin{equation}
    \hat{\chi}'=\frac{1}{k(2\pi /\beta)}\hat{\theta}', \quad \hat{t}'=\hat{t}+\hat{\chi}'
\end{equation}
As before we still identify $\hat{\theta}' \sim \hat{\theta}'  + 2\pi$ in this geometry. The new time coordinate $\hat{t}'$ corresponds to a rotating frame (in $\hat{t}',\hat{\chi}'$ plane) compared to $\hat{t}$. Now we T dualize $\hat{\chi}'$ leaving $\hat{t}'$ untouched to put the cigar geometry in the standard form\footnote{For example to compare with the formulas in section 4.1 of \cite{Kutasov:2005rr} we use following map
\begin{equation}
    \begin{aligned}
        l_s=\sqrt{2}, \quad Q=\frac{\sqrt{2}}{\sqrt{k-2}}, \quad \phi=\sqrt{2(k-2)}\hat{r}, \quad \theta=\sqrt{2(k-2)}\hat{\tau}
    \end{aligned}
\end{equation}
}
\begin{equation}\label{T2}
	\begin{aligned}
 & \Phi_C=\log(g_{s,3})-\log\( \cosh(\hat{r}) \)\\
		& ds_C^{2}=l_s^2k( d\hat{r}^2+\tanh^2(\hat{r})d\hat{\tau}^{2}+d\hat{\chi}^{2})\\
		& B_C= 0\\
	\end{aligned}
\end{equation}
In the formula above we analytically continued to the Euclidean signature and defined 
\begin{equation}
    \hat{\chi}=(2\pi/ \beta)\hat{\theta}'', \quad \hat{t}'=-i\hat{\tau}
\end{equation}
We identify
\begin{equation}
    \hat{\tau}\sim \hat{\tau}+2\pi, \quad \hat{\chi}\sim \hat{\chi}+2\pi \(\frac{2\pi}{\beta}\)
\end{equation}

In summary, we showed that the worldsheet theory of the BTZ black hole at inverse temperature $\beta$ measured in AdS units $l_{AdS}/l_s=\sqrt{k}$ is same as that of cigar geometry at temperature $1/(2\pi \sqrt{k})$ along with a decoupled compact boson of radius $r_S=(2\pi/\beta)\sqrt{k}$ measured in string units. We emphasize that the result obtained so far is only valid to the leading order in large $k$. Therefore to obtain the free energy of strings in cigar geometry we need to divide the free energy in the BTZ background by an additional factor of the length $2\pi r_S$ of the decoupled circle. Given the identification of the string coupling in a cigar with that of BTZ in (\ref{T2}), the statement above implies a non-trivial consistency test even at large $k$ when the entropy is calculated using the method in \cite{Halder:2023adw}. 
At leading order in large $k$, the entropy of the BTZ black hole is given by
\begin{equation}\label{BTZS}
    \begin{aligned}
       & S_{EBTZ} = \frac{2\pi}{g_{s,3}^2} \( \frac{2\pi }{\beta} \) \(2\pi \sqrt{k}\)
    \end{aligned}
\end{equation}
Following \cite{Halder:2023adw} and  the  T duality-related argument above
\begin{equation}
    \begin{aligned}
       & S_{cigar}=\frac{S_{BTZ}}{2\pi r_S} = \frac{2\pi}{g_{s,3}^2}=2\pi e^{-2\Phi_{C,0}}=2\pi R_C M_{C}
    \end{aligned}
\end{equation}
Here we used the map in (\ref{T2}) to express the string coupling of the cigar at the horizon $e^{\Phi_{C,0}}$ in terms of that in BTZ. The mass of the cigar $M_C$ is determined in terms of inverse temperature $2\pi R_C$ and string coupling at the tip of the cigar (see for instance \cite{Giveon:2006pr}).\footnote{
A simple way of getting the $\alpha'$ correction to the BTZ metric is considering the method in \cite{Dijkgraaf:1991ba, Tseytlin:1993df}. This method might be questionable at higher orders in $\alpha'$, but it is known to work for the leading correction of the cigar geometry. We thank J. Maldacena for a discussion of this point.
} In this sub-section for clarity we have used the subscript C for many quantities related to cigar geometry. We will drop those in subsequent discussions with the hope that this won't lead to any potential confusion for the reader.

\subsubsection{Winding condensate  description of the cigar}

According to Fateev-Zamolodchikov-Zamolodchikov duality \cite{Kazakov:2000pm, Hikida:2008pe} SL(2,R)$_k$/U(1) coset model \cite{Teschner:1997fv, Teschner:1997ft, Teschner:1999ug, Ribault:2005wp, Ribault:2005ms} is dual to Sine-Liouville theory \cite{Fukuda:2001jd, Giribet:2001ft, Giveon:2019gfk, Giribet:2021cpa}, (in the sense that some of the residues of the exact answer for the correlation functions can be obtained from the Sine-Liouville description, for the rest of the residues one has to take into account the cigar sigma model): it has two directions - radial direction is $\varphi$ and compact direction $\tau \sim \tau+2\pi R, R=\sqrt{k}$. The radial direction has a background charge $Q=1/\sqrt{k-2}$, giving central charge $$c=1+1+6 Q^2=\frac{3k}{k-2}-1$$ The Sine-Liouville Lagrangian is given by\footnote{Compare this with the asymptotic geometry of the cigar in (\ref{cigarS}) with the identification 
\begin{equation}
    \varphi=-\sqrt{k-2}\hat{r}, \quad \tau=\sqrt{k} \hat{\tau}
\end{equation}
}
\begin{equation}\label{SineLiouvilleAction}
\begin{aligned}
    & S=\frac{1}{2 \pi}\int 2 d^2 \sigma (\partial \varphi \bar{\partial} \varphi+\frac{1}{4b'}\varphi R+\partial \tau \bar{\partial} \tau +\frac{\pi\mu}{2b^{'2}} (W^{+}+ W^{-}))\\ 
  & W^{\pm}:= e^{\mp i\sqrt{k}  (\int_{(0,0)}^{(z,\bar{z})}\partial \tau dz'- \int_{(0,0)}^{(z,\bar{z})}\bar{\partial}\tau d\bar{z}')} \ e^{\sqrt{k-2}\varphi},~~ \mu:=2b^{'4} \bigg(\frac{\mu'\gamma(b^{'2})}{\pi }\bigg)^{1/2},\\
  & b':=\frac{1}{b''}, ~~ \pi \mu'\gamma(b^{'2}):=(\pi \mu''\gamma(b^{''2}))^{1/(b^{''2})}, \quad  b''=\frac{1}{\sqrt{k-2}}, ~~ \mu''=\frac{b^{''2}}{\pi^2}
\end{aligned}
\end{equation}

The precise formula for the cosmological constant is obtained from a similar discussion in AdS$_3$, consult \cite{Halder:2022ykw} for more details. Conceptually one can think about it as follows: first one chooses a specific coordinate system in the bulk and a set of canonical fields. With this choice the expression for the string coupling at the horizon is a specific function of the bulk fields (acting with a field redefinition would change this function and from now on we don't allow that). Now  one builds the non-linear sigma model worldsheet theory with $\alpha'$ corrections as in (\ref{cigarS}). On the Sine-Liouville side (\ref{SineLiouvilleAction}) string coupling is given by
\begin{equation}\label{gsCigar}
    g_{s,C}=e^{\Phi_\infty}
\end{equation}
Using Sine-Liouville theory we demand to reproduce the string amplitudes of the sigma model side including $\alpha'$ corrections. This procedure and formal consistency requirements as a conformal field theory fixes the cosmological constant as above. 

In the free theory, we are working with the usual conventions\footnote{The T dual coordinates are given by
\begin{equation}
     \int_{(0,0)}^{(z,\bar{z})}\partial \tau dz'- \int_{(0,0)}^{(z,\bar{z})}\bar{\partial}\tau d\bar{z}'= \tau_L(z)-\tau_R(\bar{z})
\end{equation}
The lower limit of the integration would not matter for the correlation function of properly quantized vertex operators and we have dropped in on RHS.
} 
\begin{equation}\label{OPE}
	\begin{aligned}
        &\tau(z,\bar{z})=\tau_L(z)+\tau_R(\bar{z}), \quad  \tau_L(z)\tau_L(0)\sim -\frac{1}{2}\ln z
	\end{aligned}
\end{equation}
etc. The properly quantized vertex operators take the following form\footnote{In the spirit of the T duality discussed in the previous section, one can ask the relation between the vertex operators of the AdS$_3$ sigma model and the cigar geometry. The local vertex operators of SL(2,R) WZW model  can be thought of as a product of $T_{j,m,\bar{m}}$ along with those of a time-like compact boson (we are working in Lorentzian signature for AdS$_3$ while keeping the cigar geometry Euclidean, therefore the additional compact boson is timelike) of radius $\sqrt{k}$. Indeed the holomorphic dimension of the vertex operator in  AdS$_3$ can be written as a sum of the cigar contribution and that of the compact boson \cite{Giveon:1999px}
\begin{equation}
    -\frac{j(j+1)}{k-2}+\frac{m^2}{k}-\frac{1}{k}(m-\frac{k\nu}{2})^2
\end{equation}
Here $m-\frac{k\nu}{2}$ is the momentum of the compact boson.
}
\begin{equation}\label{Vop}
    \begin{aligned}
       & T_{j,m,\bar{m}}=e^{i\frac{2m}{\sqrt{k}}\tau_L-i\frac{2\bar{m}}{\sqrt{k}}\tau_R-\frac{2}{\sqrt{k-2}}j (\varphi_L+\varphi_R)} \quad j=-\frac{1}{2}+i\mathbb{R}\\
       & m=\frac{p-k\nu}{2}, \quad \bar{m}=-\frac{p+k\nu}{2}, \quad p,\nu \in \mathbb{Z}\\
       & h= -\frac{j(j+1)}{k-2}+\frac{m^2}{k}, \quad \bar{h}= -\frac{j(j+1)}{k-2}+\frac{\bar{m}^2}{k}
    \end{aligned}
\end{equation}
Here $p,\nu$ are momentum and winding number around $\tau$ circle and $h,\bar{h}$ are the holomorphic and anti-holomorphic dimension.

\subsubsection{Stringy area operator}

The goal of this section is to calculate the leading order at small string coupling thermal entropy of string theory in the cigar background.
We will proceed to evaluate the partition function of the worldsheet CFT on the unit sphere\footnote{For similar discussion in the simpler context of Liouville theory see \cite{Mahajan:2021nsd}.}
\begin{equation}
    \begin{aligned}
&  Z_{CFT}=  Z^{SL}  Z_c
    \end{aligned}
\end{equation}
$Z_c$ is the contribution of the time-like Liouville theory and ghosts that are independent of temperature. For the BTZ black hole, we have (compare the formula in \cite{Halder:2023adw} with   (\ref{BTZS})) 
\begin{equation}\label{zc}
    Z_c=\frac{\pi^3}{2}
\end{equation}
Based on the T duality argument we expect the same factor for the cigar geometry. 
The partition function of the cigar CFT is given by
\begin{equation}
    Z^{SL}= Z^{SL}_0 C^{SL}_{S^2}
\end{equation}

The contribution of zero modes on the cigar circle $Z^{SL}_0$ is much more subtle.
In the example of BTZ black hole, it was pointed out in \cite{Halder:2023adw} that the right contribution is obtained by: first using a diffeomorphism to put the metric in a form where the length of the thermal circle $\hat{\xi}=i\hat{t}$ is independent of the temperature of the black hole and that amounts to a space circle of length proportional to the temperature in AdS units. Now we turn to explain this factor in the frame $\hat{\theta} \sim \hat{\theta}  + 2\pi (2\pi/\beta), \hat{\xi} \sim \hat{\xi}  +2\pi$ with
\begin{equation}\label{EBTZfields}
	\begin{aligned}
 & \Phi=\log(g_{s,3})\\
		& ds^2=l_{AdS}^2(d\hat{r}^2+ \cosh^2(\hat{r})d\hat{\theta}^2+\sinh^2(\hat{r})d\hat{\xi}^2)\\
		& B=-i  l_{AdS}^2\sinh^2 (\hat{r})  \ (d\hat{\theta} \otimes d \hat{\xi}-d\hat{\xi} \otimes d\hat{\theta})\\
	\end{aligned}
\end{equation}
The contribution of the boundary torus zero modes can be written as
\begin{equation}
    2\pi \(\frac{4\pi^2}{\beta}\)=  \( \frac{2\pi l_{AdS} (2\pi/\beta)e^{\hat{r}}}{l_s}\) \(\frac{2\pi R' }{l_s}\), \quad R' =\frac{l_s^2}{l_{AdS}e^{\hat{r}}}
\end{equation}
On RHS the first factor in the bracket comes from the length of the spatial circle in string units at the AdS boundary, and the second factor comes from the length of the T-dual Euclidean time circle at the AdS boundary, again measured in string units. In the case of the two-dimensional cigar, we do not have a transverse space circle, however, we expect the contribution to be proportional to the temperature of the cigar in string units following the logic above. More precisely, we are using the winding condensate description and we are supposed to evaluate the partition function after performing the T duality on the Euclidean time circle. The correlation function of the winding condensate measured in units of the partition function of the free Linear dilaton coupled to the Euclidean circle is invariant under the T duality, and the zero mode factor we are after is the partition function of this free theory with the integration over the radial constant mode omitted (and non-zero modes on a spherical worldsheet with constant curvature contributes a constant factor). Therefore the required zero mode factor is the length of the T-dual Euclidean time circle at the asymptotic infinity that can be easily read off from (\ref{cigarS}) to be\footnote{Note that the partition function of a compact boson of radius $R$ on genus $g$ Rieman surface goes like $R^{1-g}$ and under $T$ duality the correlation functions measured with respect to the partition function are invariant, however, the partition function is not. This is the reason during T duality we shift the dilaton to $\Phi \to \Phi -\log(R)$ (see (\ref{Tduality})) to make sure the string theory based on the compact boson is unchanged under the T duality. } 
\begin{equation}\label{zSL}
    Z^{SL}_0=\frac{2\pi}{\sqrt{k}} =2\pi \(\frac{4\pi^2}{\beta}\)\frac{1}{2\pi r_S}
\end{equation}
This is in complete agreement with what we expect based on the T-duality arguments of the previous sub-section. 
Integrating out the radial constant mode and performing the Wick contraction gives (for more details of this standard process see \cite{Halder:2022ykw})
\begin{equation}\label{res}
	\begin{aligned}
		C_{S^2}^{SL}=& \frac{2\pi}{b'}\Gamma(-2s')  \bigg( \frac{-\mu}{2b^{'2}}\bigg)^{2s'}\frac{\Gamma(2s'+1)}{\Gamma(s'+1)^2}\int\prod_{i=1}^{s'}  d^2z_i \int \prod_{i'=1}^{s'}  d^2z'_{i'} \\
  &~~~~~~~~~~~~~~~~~~~~~~\prod_{i'<j'}   \bigg[ (z_{i'j'})  (\bar{z}_{i'j'})\bigg] \ \prod_{i<j}   \bigg[ (z_{ij})  (\bar{z}_{ij})\bigg] ~~ \prod_{i,j'}   \bigg[ (z_{ij'})^{1-k}  (\bar{z}_{ij'})^{1-k}\bigg]
\end{aligned}
\end{equation}
Here we have defined $s'=1/b^{'2}$. This formula is valid for integer $s'$. Using conformal maps it can be shown that the expression above has an infinite factor of the volume of the unfixed gauge group PSL(2,$\mathbb{C}$).\footnote{For a discussion in of flatspace string theory see \cite{Erbin:2019uiz} (see also \cite{Eberhardt:2021ynh, Eberhardt:2023lwd}). } One can calculate the finite prefactor by essentially fixing location of three of the winding operators.\footnote{Similar calculation in the context of time-like Liouville theory is available in \cite{Giribet_2012}.}
The physics of the expression above is related folded strings originating from $W^+ W^-$ OPE \cite{itzhaki2024stringssurprise} (see also  \cite{Giveon:2016dxe,  Itzhaki:2018glf, Giveon:2019gfk, Giveon:2020xxh}). 

Next we analytically continue this formula suitably to get the answer for generic values of $k$. The calculation for this part is essentially same as the one in \cite{Halder:2023adw} and we borrow the results from there
\begin{equation}
    \begin{aligned}
      C_{S^2}^{SL}= \frac{1}{\Gamma(-1)} \frac{1}{\pi^3} \bigg(\sqrt{k-2}-\frac{1}{\sqrt{k-2}}\bigg) z_{PSL(2,\mathbb{C})} 
    \end{aligned}
\end{equation}
\textit{We would like to emphasize that this result takes into account the effect of the winding condensate on metric and dilaton (for more details see \cite{Halder:2023adw}).}
The volume of PSL(2,$\mathbb{C}$) will drop out of the calculation once we consider string theoretic free energy. However, the overall factor of the Gamma function makes the contribution zero. This issue is related to the lack of understanding the worldsheet theory in presence of a spacetime boundary (a discussion of this in the context of string field theory appears in \cite{Erler:2022agw, Cho:2023khj})\footnote{Also see the related topics in \cite{Ahmadain:2022tew, Ahmadain:2022eso}.}. It was shown in  \cite{Halder:2023adw} that the correct finite result in $\alpha'\to 0$ limit can be obtained by performing a version of the target space replica trick \cite{Lewkowycz:2013nqa} on the worldsheet (it was shown that the correct entropy can be obtained by introducing a conical singularity in the target space through infinitesimally non-integer winding condensate on the worldsheet and finally differentiating the result with respect to the infinitesimal parameter).  In this limit, the slope of the linear dilaton in the bulk goes to zero. As a result, it was not important to understand how to change the slope of the dilaton when we introduced the conical singularity for the replica background. In this work, we are after stringy corrections to thermal entropy and therefore it is absolutely important that we fix this ambiguity. Hints come from the discussion of similar issues in the target space \cite{Gibbons:1992rh, Kazakov:2001pj}. The important idea is that first we choose a cut-off surface in the radial direction at $\hat{r}=\hat{r}_c$ and fix the string coupling $\Phi(\hat{r}_c)$ on the cut-off surface while varying the temperature
\begin{equation}
    k \to k(1-2\delta), \quad \delta \to 0+
\end{equation}
In other words, we change the string coupling at the tip of the cigar $\Phi_0$ while varying the slope of the dilaton in a coordinated way such that the  $\Phi(\hat{r}_c)$ remains fixed. Finally, we identify the resulting background with the periodicity of the fields as if the temperature and the gradient of the dilaton are not changed at all to create the conical singularity at the tip of the cigar. As a summary, the conical singular replica background is created by the following replacement in the calculation of the previous section
\begin{equation}
    \begin{aligned}
        W^\pm \to W^{\pm}_{\delta}=  e^{\mp i\sqrt{k} (1-\delta) (\int_{(0,0)}^{(z,\bar{z})}\partial \tau dz'- \int_{(0,0)}^{(z,\bar{z})}\bar{\partial}\tau d\bar{z}')} \ e^{\sqrt{k-2}(1-\frac{1}{1-\frac{2}{k}}\delta)\varphi}
    \end{aligned}
\end{equation}
Such a replacement in general creates a reflected wave, and the prescription of  \cite{Halder:2023adw} is not to include that in the path integral calculation.
The modified winding operator is not $(1,1)$ in the (undeformed) free theory and it is infinitesimally off-shell everywhere. It carries winding $1-\delta$ around the Euclidean time circle. 
The modified result to the first order in $\delta$ is 
\begin{equation}
	\begin{aligned}
			 C_{S^2,\delta}^{SL}=& z_{PSL(2,C)}\frac{2\pi}{b'}\Gamma(-2s')  \bigg( \frac{-\mu}{2b^{'2}}\bigg)^{2s'}\frac{\Gamma(2s'+1)}{\Gamma(s'+1)^2} \prod_{I=3}^{s'}  d^2z_I \int \prod_{I'=2}^{s'}  d^2z'_{I'}~~\\ & \bigg[ (\prod_I |z_{I}|^{2(1+(\dots) \delta)}|z_{I}-1|^2)(\prod_{I'} |z'_{I'}|^{2(1-k+ a(k)\delta)}|z'_{I'}-1|^{2(1-k)} )\bigg] \\
		& ~~~~~~\prod_{I'<J'}   \bigg[ |z_{I'J'}|^2 \bigg] \ \prod_{I<J}   \bigg[ |z_{IJ}|^2  \bigg] ~~ \prod_{I,J'}   \bigg[ |z_{IJ'}|^{2(1-k)} \bigg]
	\end{aligned}
\end{equation}
In the notation of \cite{Halder:2023adw} replica factor
\begin{equation}\label{replicaF}
    \begin{aligned}
        a(k)=\frac{2k}{k-2}
    \end{aligned}
\end{equation}
In the process of the path-integral, we have automatically taken the cut-off surface to infinity. While taking the derivative with respect to $\delta$ to produce the thermal entropy it is important that we hold the asymptotic string coupling fixed.  The thermal entropy as computed from the replica trick is given by\footnote{The subscript `NH' reminds us that the contribution is from the strings living near the horizon. }
\begin{equation}\label{replicaEE}
    S_{NH}=\lim_{\delta\to 0}\(-\partial_{n}(\log Z(n)-n\log Z(1)) \bigg|_{n=1+\delta}\), \quad \log Z=\frac{1}{g_{s,C}^2} \frac{Z_{CFT}}{ z_{PSL(2,\mathbb{C})}}
\end{equation}
\begin{figure}[ht]
	   \centering
	   \includegraphics[width=0.8\textwidth]{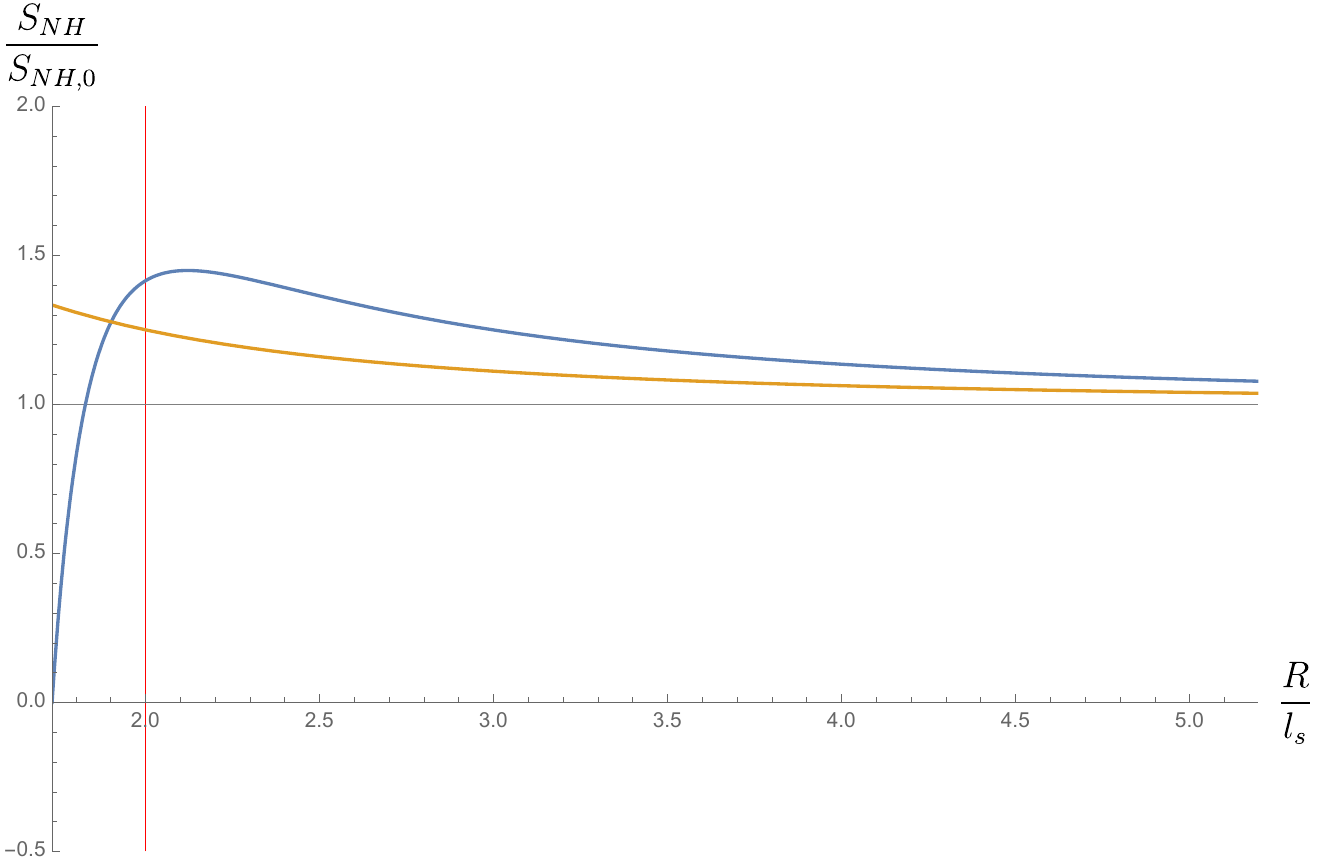}
\caption{In blue, we are plotting the contribution $S_{NH}$ of the near horizon cigar region to the entropy of the Schwarzschild black hole in large $D$ (the relevant formula is available in (\ref{finalEntropy})). The gray line and the yellow curve are respectively the leading order Bekenstein-Hawking entropy $S_{NH,0}$ and the Wald entropy including just the first stringy correction. The red line represents the Hagedorn radius $R_{H}=2l_s$ of the empty flat space and the origin corresponds to the Hagedorn radius $R_{C,H}=\sqrt{3}l_s$ in linear dilaton background of the cigar geometry. }
 \label{fig2}
\end{figure}
$\log Z$ is the string theoretic free energy. The entropy above can be obtained as string theoretic  one-point function of the non-local area operator 
\begin{equation}\label{areaOp}
  A=  \frac{\mu}{b^{'4}} \int  d^2 \sigma \frac{\partial}{\partial \delta} (W_\delta^{+}+ W_\delta^{-})\bigg|_{\delta=0}
\end{equation}
The reason the one point function of $A$ is non-zero lies in the fact that its correlation functions with legitimate vertex operators in the BRST cohomology will have a branch cut. One can also consider the one point function of $A$ on a toroidal worldsheet, which is expected to produce the quantum corrected area in the one loop contribution to the thermal entropy, while the rest of the result should come from the entanglement entropy of the bulk fields outside the horizon. In our formalism, we avoided such issues by considering the exponentiated version of the operator. It is expected that the $\alpha' \to 0$ limit of the operator $A$ is intimately related to the algebra of observables in the target space \cite{Leutheusser:2021qhd, Leutheusser:2021frk, Witten:2021unn, Chandrasekaran:2022eqq}.
The entropy can be easily calculated to be
\begin{equation}
\begin{aligned}
    S_{NH}=&\Gamma(-1) \ a(k) \( \frac{1}{g^2_{s,C}} Z_c   \ Z_0^{SL} \) \frac{C_{S^2}^{SL}}{z_{PSL(2,\mathbb{C})}}   \\
    = & \(\frac{2\pi }{g_{s,C}^2 }   \)  \sqrt{\frac{k}{k-2}}  \(1-\frac{1}{k-2}\)\\
    = & \(\frac{2\pi A_{D-2} r_0^{D-2}}{g_s^2 }   \)   \frac{k}{k-2}  \sqrt{\frac{k}{k-2}}  \(1-\frac{1}{k-2}\)\\
\end{aligned}
\end{equation}
We used (\ref{zc}), (\ref{zSL}), and (\ref{gsCigar})  to obtain the expression above. For convenience, we express the final result for the entropy coming from the near-horizon cigar region as
\begin{equation}\label{finalEntropy}
\begin{aligned}
    & S_{NH}=S_{NH,0}  \frac{(1-3 (2\pi T)^2)}{\left(1-2 (2\pi T)^2\right)^{\frac{5}{2}}} \\
    & S_{NH,0}=a_{D-2}\(\frac{1}{(2\pi T)^2}-2\)^{\frac{D-2}{2}}, \quad  a_{D-2}=\frac{A_{D-2} }{4G_D}\(\frac{D}{2}\)^{D-2}
\end{aligned}
\end{equation}
The mass of the cigar geometry is determined by thermodynamics to be
\begin{equation}\label{finalMass}
    M_{NH}=\int_{\frac{1}{2\pi \sqrt{3}}}^T  \(\frac{\partial S_{NH}}{\partial T}\)T dT
\end{equation}
The explicit functional form is a little complicated and we won't mention it here. We will come back to the specific limits of this formula later. 

\subsection{Analyzing the stringy entropy formula}

\subsubsection{The black hole limit}

We expect to recover the thermodynamics of the black hole in the low-temperature limit. This is easily seen from (\ref{finalEntropy}) as follows
\begin{equation}\label{BHlimit}
\begin{aligned}
    S_{NH}
    = &   \(\frac{A_{D-2} r_0^{D-2}}{4G_D }   \) \(1+\frac{2}{k}+\frac{5}{2k^2}+O\(\frac{1}{k^4}\) \)
\end{aligned}
\end{equation}
 We get the precise coefficient for the first $\alpha'$ correction to the entropy as given in  (\ref{largeDth}) to the leading order in large $D$ using the map (\ref{level}). Before taking the replica factor $a(k)$ into account the leading order stringy correction to the entropy vanishes, whereas after taking the correction into account we get the positive sign as expected from the large $D$ analysis. Therefore the stringy corrections from the replica trick play a very important role in reproducing the large $D$ answer even qualitatively. Since both the first and second stringy correction to entropy is positive, the stringy correction to the entropy keeps growing as we increase temperature, i.e., decrease $k$, until we are very close to the Hagedorn temperature of flat space at $k=4$. The entropy reaches a local maximum at
 \begin{equation}
     T=\frac{1}{3\sqrt{2}\pi}\approx\frac{1}{2\pi R_H}, \quad R_H=2
 \end{equation}
As we approach the inverse Hagedorn radius $R \to R_H=2$ of the asymptotic empty flat space, a winding condensate in the Euclidean time circle develops in the asymptotic region and its contribution to entropy becomes important. Therefore in this limit, the entropy of the black hole is composed of not only the contribution of the near horizon cigar region that we calculated in this paper but also of the winding mode in the far away flat space. In other words, the entropy of the black hole is not well-represented by the blue curve in figure \ref{fig2} as we increase temperature and reach near the red line. 
In asymptotically flat space in large dimensions $D$, any normalizable solution of the Horowitz-Polchinski type near Hagedorn temperature would be necessarily stringy \cite{Chen:2021dsw}. Therefore it is challenging to reproduce the behaviour of entropy from a target space gravity analysis in large $D$ directly.

\subsubsection{The winding condensate limit}

On the other hand, from our formula (\ref{finalEntropy}), we see that as we approach  $R\to R_{C,H}=\sqrt{3}$ the $g_{s,C}^{-2}$ contribution to the entropy of the near horizon cigar black hole vanishes  (this does not say anything about the large $D$ black hole of course)\footnote{It is interesting to compare this result against the phase diagram in \cite{Betzios:2022pji} for small $g_s$ and $R<2$. We thank Xi Yin for a related discussion. }. This is precisely when the Euclidean time winding condensate in the cigar background becomes non-normalizable and massless in empty linear dilaton background \cite{Kazakov:2000pm, Karczmarek:2004bw, Maldacena:2005hi} and we expect the black hole to turn into a gas of free strigs \cite{Kutasov:2005rr, Giveon:2005mi, Giveon:2006pr}. The picture is similar to that of a small black hole in AdS - when the temperature is near the Hagedorn temperature of the thermal AdS string theory the small black hole first turns into interacting strings in thermal AdS governed by the Horowitz-Polchinski-like solution and then at Hagedorn temperature it turns to free strings \cite{Martinec:2021vpk, Halder:2023nlp, Agia:2023skp, Urbach:2022xzw, Urbach:2023npi}. Our worldsheet analysis takes the effect of the string-corrected complicated profile of the dilaton and metric as given in (\ref{cigarS}) into account while studying the winding condensate on the Euclidean time circle and predicts a very simple formula for the entropy in this limit
\begin{equation}
    \begin{aligned}
       & S_{NH}=2\pi R_{C,H} M_{NH}+\frac{\xi}{2}M_{NH}^2+O(M_{NH}^3), \quad \xi=\frac{8\pi^2}{9R_{C,H}a_{D-2}}\\
       & M_{NH}=\frac{3R_{C,H}}{\pi}a_{D-2}(R-R_{C,H})+\frac{9}{\pi}D a_{D-2}(R-R_{C,H})^2+O((R-R_{C,H})^3)
    \end{aligned}
\end{equation}
The respective scaling of mass and entropy looks very close to the general analysis of Horowitz-Polchinski type solutions in flat space given in \cite{Chen:2021dsw} extrapolated to three spacetime dimensions. This suggests thinking of the cigar in terms of coset of AdS$_3$ might be very helpful to understand this limit - especially in terms of the weakly coupled boundary theory (along the lines of \cite{Aharony:2005ew}). 

Discussion of this limit $R\to R_{C,H}=\sqrt{3}$ also appears in \cite{Chen:2021emg} treating the winding condensate as a probe around linear dilaton spacetime. It is not clear if the conclusions there hold once all stringy backreactions are taken into account.\footnote{For some discussion of the back-reaction see \cite{Brustein:2021qkj}.} We emphasize that our conclusion is only valid to the leading order in $g_s$, and our results break down when
\begin{equation}
    a_{D-2}(R-R_{C,H})\sim 1
\end{equation}
One has to take into account higher order in $g_s$ effects for higher temperatures. It is possible that once we look into quantum string theory the conclusions of \cite{Chen:2021emg} remain valid.

\section{Quantum strings in large D and future directions}\label{sec4}

So far in this paper, we have discussed classical string theory of the near horizon geometry and the timelike Liouville theory (tLT) played the role of a spectator for the gravitational dynamics in $D$ spacetime dimensions.\footnote{We have focussed on the thermodynamics, however, it should be possible to compare for instance two-point function of cigar vertex operators with that of the flat space and see the physics of the quasi-normal modes. We leave such exercise for the future.} However one can also excite the degrees of freedom of the timelike Liouville theory, which will change the physics completely - now the strings are moving in effectively $D+1$ dimensions whose spatial section is the Euclidean $D$ dimensional black hole and the role of time is played by the time-like Liouville theory. The proper target space interpretation of this theory is unknown. In this section, we will take a look at these questions and point out a particular simplification at ultra-low temperatures. 

Consider the $n$ point correlation function of on-shell vertex operators of the following form 
\begin{equation}
    \mathcal{V}_{j}^\pm = \(\prod_{i=1}^n \frac{\Gamma(\pm\frac{k}{2}-j)}{\Gamma(1\mp\frac{k}{2}+j)}\) T_{j,\mp \frac{k}{2},\mp \frac{k}{2}}\hat{T}_{\hat{\alpha}}, \quad \nu=\pm1, \quad p=0
\end{equation}
where $T_{j,m,\bar{m}}$ is given in (\ref{Vop}) and $\hat{T}_{\hat{\alpha}}$ is the vertex operator of the time-like Liouville theory. The condition of being $(1,1)$ fixes $\hat{\alpha}$ entirely in terms of $j$. We choose the first  $n-1$ operators of winding $\nu=+1$ and the $n$th operator with $\nu=-1$. Our central observation is that the $n-$point correlation function then takes the following form (this can be easily seen from small generalization of the calculation done in \cite{Halder:2022ykw})\footnote{The proportionality constant depends on $n,k$.}
\begin{equation}
    \begin{aligned}
        &\langle  \mathcal{V}^-_{j_n}(\infty) \prod_{i=1}^{n-1} \mathcal{V}^+_{j_i}(z_i,\bar{z}_i) \rangle \propto \langle T_{\alpha_n}(\infty) \prod_{i=1}^{n-1} T_{\alpha_i}(z_i,\bar{z}_i) \rangle_{\tilde{\mu},\tilde{b}} \langle \hat{T}_{\hat{\alpha}_n}(\infty) \prod_{i=1}^{n-1} \hat{T}_{\hat{\alpha}_i}(z_i,\bar{z}_i) \rangle_{\hat{\mu},\hat{b}}
    \end{aligned}
\end{equation}
On the right-hand side, $T_\alpha$ is a vertex operator in an effective spacelike Liouville theory (sLT) with parameter $(\tilde{\mu}=\mu'',\tilde{b}= b'')$ (our conventions of spacelike Liouville theory is same as that of \cite{Halder:2022ykw}) and we have conveniently parameterized 
\begin{equation}
    j=-\frac{1}{2}-i\sqrt{k-2}P, \quad \alpha =\frac{1}{2}(\sqrt{k-2}+\frac{1}{\sqrt{k-2}})+iP, \quad P\in \mathbb{R}
\end{equation}
This parameterization ensures the dimension of $T_{j,\mp \frac{k}{2},\mp \frac{k}{2}}$ is the same as that of the spacelike Liouville operator $T_\alpha$. 

According to Zamolodchikov \cite{Zamolodchikov:2005fy}, such products will simplify dramatically if
\begin{equation}
    c_{\text{sLT}}+c_{\text{tLT}}=26 \implies 1+6\(b''+\frac{1}{b''}\)^2+26-D=26 \implies  k=\frac{D-1}{6}+O\(\frac{1}{D}\)
\end{equation}
We will call this limit ultra-low temperature limit. In this limit, not only the genus zero correlation function but all higher genus correlation functions are expected to simplify\footnote{A different type of simplification for quantum gravity in large $D$ is speculated in \cite{Strominger:1981jg}. } - this is mathematically similar to the Virasoro minimal strings discussed in \cite{Collier:2023cyw}.\footnote{See similar discussions in \cite{Goel:2023svz, Narovlansky:2023lfz}.} Given that Virasoro minimal strings are dual to a double-scaled matrix model, we expect a matrix model dual for this sector of correlation functions in the ultra-low temperature limit. Finally, one can extrapolate the results to $D=26$ (this relies on existance of a universal zero central charge limit of time-like Liouville-like theories) and see if we can learn any non-perturbative properties of the Schwarzschild black hole in large $D$\footnote{The extrapolated value for the level $k = 4+1/6$ is actually close to the Hagedorn value of the empty flatspace $k=4$. }. We leave this fascinating question for future exploration.

\section{Acknowledgement}

We thank  David Gross, Zohar Komargodski, Juan Maldacena for insightful  comments. Also we thank Yiming Chen, Scott Collier, Daniel Harlow,  David Kutasov, Shiraz Minwalla, Mukund Rangamani, Herman Verlinde, and Xi Yin for the helpful discussions. We thank Joydeep Naskar for discussing similar projects. The work of DLJ and IH is supported in part by DOE grants DE-SC0007870 and DE-SC0021013.

\appendix

\providecommand{\href}[2]{#2}\begingroup\raggedright\endgroup

\end{document}